\documentclass[10pt]{article}

    \PassOptionsToPackage{numbers, compress}{natbib}

\usepackage[final]{neurips_2020}

\usepackage[utf8]{inputenc} 
\usepackage[T1]{fontenc}    
\usepackage{hyperref}       
\usepackage{url}            
\usepackage{booktabs}       
\usepackage{amsfonts}       
\usepackage{amsmath}        
\usepackage{nicefrac}       
\usepackage{microtype}      
\usepackage{multirow}
\usepackage{xcolor}
\usepackage{tocloft}
\usepackage[inline, shortlabels]{enumitem}  

\title{Convening during COVID-19: Lessons learnt from organizing virtual workshops in 2020}
\date{August 2020}
\author{%
 Mandana Samiei‡\textnormal{\textsuperscript{1}}\quad
 Caroline Weis‡\textnormal{\textsuperscript{2}}\quad
 Larissa Schiavo‡\textnormal{\textsuperscript{3}}\quad\\
 Tatjana Chavdarova\textnormal{\textsuperscript{4}}
 Fariba Yousefi\textnormal{\textsuperscript{5}}
 \thanks{
 \textsuperscript{1}\,McGill\,University %
 \textsuperscript{2}\,ETH\,Zurich %
 \textsuperscript{3}\,formerly\,Open\,AI 
 \textsuperscript{4}\,EPFL %
 \textsuperscript{5}\,University\,of\,Sheffield %
\texttt{mandana.samiei@mail.mcgill.ca, caroline.weis@bsse.ethz.ch, larissafschiavo@gmail.com }}\\[1em]
 
 \textbf{‡These authors contributed equally to this manuscript.}
}
\begin{document}

\maketitle
\tableofcontents

\newpage
\section{Overview}
This year, for the first time ever, Women in Machine Learning (WiML) decided to host an ``un-workshop'' as its small event at ICML, instead of its traditional meal-based event.  Un-workshops focus on \textit{participant-driven structured discussions} on pre-selected topics. This format encourages participant interactions, emphasizes attendee contributions, and provides opportunities for future collaborations (see~\S~\ref{sec:un-conf}). This event became a ICML 2020 Affinity Workshop held on Monday, July 13, 2020.  Due to the global COVID-19 pandemic, and ICML going virtual, the ``WiML Un-Workshop'' also went virtual.  
 
 When confronted with the many challenges, restrictions, but also opportunities associated with virtual events, many critical decisions have to be made with limited data. In this manuscript, we -- the organizers of the first WiML Un-Workshop -- share our experiences with the hope that it will help future event organizers host successful virtual events under similar conditions. For completeness, we list our main steps we took and decision-making processes, some of which may apply to physical conferences too. 

\subsection{Background}
 
This report is an account of the authors' experiences as organizers of the WiML ``Un-Workshop'' event at ICML 2020.  For clarity, this event was different from the ``WiML Workshop'', which is usually co-located with NeurIPS.  
Women in Machine Learning (WiML)'s mission is creating connections within a small community of women working in machine learning, in order to encourage mentorship, networking, and interchange of ideas and increase the impact of women in the community. More information can be found on the official WiML website (https://wimlworkshop.org/mission/).

\subsection{Un-conferences}\label{sec:un-conf}
The ``un-workshop'' is based on the concept of an un-conference. 
Un-conferences aspire to increase participants' engagement by reducing formal elements of a conference, such as talks and panel discussion, and encouraging informal discussions between participants~\citep{socmlgoodfellow}.
The format gained more traction in the machine learning community with the first Self-Organizing Conference on Machine Learning (SOCML), first organized in 2016, and repeated installments in 2017 and 2018~\citep{socmlgoodfellow}.

WiML chose the un-conference format for its first ever virtual workshop after the pandemic started. 
In virtual form, the format of a traditional conference bears the risk of increased detachment between speaker and audience, and an impersonal atmosphere. 
With four invited keynote speakers and 27 breakout sessions organized and led by participants throughout the day~\citep{wiml2020program}, 
a large focus was put on small-group discussion via video-conference chats.
In our experience this format allowed for un-workshop participants with different scientific and cultural backgrounds to meet over the shared interest in a specific breakout session topic, leading to a high-energy atmosphere with lively discussions.

\section{Registration and participation}

\subsection{Call for Participation (CfP)}
Registration and call for participation were carried out using a Google form designed by the organizers. 
This form should be designed very carefully, as it will initiate the first communication between participants and organizers. Failing to ask for necessary information in this step can make follow-up calls and emails necessary. Moreover, ambiguously-phrased questions may lead to a flood of emails asking for clarifications. With this in mind, we did several rounds of redesigning this Google form. Organizers tested the form while assuming the role of every level of participant (attending only, poster presenter, etc.) to ensure that the consecutive questions are aligned even if parts of the form were skipped.
\newpage
After deciding on a rough format (invited talks, breakout sessions, and poster session), we designed a form which contains the following sections: 

\begin{enumerate}
    \item Personal information 
    \item Breakout session proposal and leaders information
    \item Poster 
    \item Facilitators and volunteers 
    \item Registration fee funding 
    \item Elaboration on status (exceptional circumstances, early career researcher, etc.) 
\end{enumerate}
We asked all breakout session leaders and poster presenters to specify their area of interests and this helped us to organize the sessions and talks along with the program. 
We prepared the Call for Participation (CfP) and the registration Google form almost in parallel. At this point the event was 8 weeks away. In the CfP, we introduced various ways of participating in the un-workshop, presenting posters, leading a breakout session, moderating sessions, and volunteering for technical issues.

\subsection{Sign-up}

The sign-up form was based on given templates from previous WiML workshops. Due to the significant change in format and the virtual nature of the event, it was difficult to predict the number of attendees based on past events. In the initial phase, it was important to foresee different outcomes, such as the requirement of volunteers, so as to include such questions in our sign-up form.  
%

%
The form also asked for participants to indicate a willingness to help as a facilitator or a volunteer, to have a sufficient number of potential volunteers to contact later on.
Due to worldwide participation, we also collected information on the preferred timezone of each participant, which  was later  used to decide on the time and the span of the program, see \S~\ref{sec:time-span}.

\subsection{Participant submissions}
Participants were offered two ways of submitting their own work/ideas to the un-workshop, proposals for a breakout session, or submitting a poster. 

\paragraph{Breakout session proposal.}
A breakout session is a one hour free-form discussion overseen by 1-3 leaders and with assistance from 1-2 facilitators to take notes and encourage participant interactions.
The submission included a one-page blind proposal for which we provided a template in the CfP form. The names and the biographies of leaders were submitted separately in the application form.

\paragraph{Poster submission.}
We accepted submission either as an abstract in free text or posters in PDF through the Google Form (see Section~\ref{platformsparagraph}).
As the format was virtual and there was no physical limit to the number of posters, we wanted to give everyone who wanted to present a poster the opportunity to do so. 
To review the large number of submissions, in hindsight it would have been easier to settle on one submission file format to find the content of each poster submission easily.
To speed up the review process, we recommend that organizers only ask for an early abstract submission in the application form. Once the acceptance notification was released, a final PDF version will be asked closer to the day of the event.

\paragraph{CMT, EasyChair vs. Google Form.}
We used a Google Form since CMT did not support PNGs at that time, however PNGs are the only applicable format for posters in the Gather.town platform. For the future, we strongly encourage Gather.town to allow for files in PDF format in their interface. This way, organizers can consider CMT or EasyChair over Google Form while in the first two platforms it is more convenient to organize and evaluate proposals and posters. 

\section{Program}\label{sec:program}

In this section, we will briefly outline how we decided on our un-workshop program and discuss things needed to be considered.
We chose a completely different format of WiML workshop by introducing the concept of breakout sessions. We minimized the number of talks to 4 and increased the number of breakout sessions to 27 to encourage a dynamic spirit during the whole event. This was a great opportunity for junior and early-career researchers to communicate and network with leading people in the field. 
\subsection{Program items}
\begin{itemize}
    \item Invited Talks
    \item Breakout Sessions 
    \item Affinity Group Joint Poster Session 
    \item Panel
    \item Socials (Welcome session and coffee breaks)
\end{itemize}
\subsection{Timezones}\label{sec:time-span}
 Our goal was to provide an interesting program to participants from all timezones, thus we needed relevant program points, such as invited speakers, in the morning and in the evening. From the sign-up form, we knew in advance to expect ~70-80\% of un-workshop participants to be based in North America and Europe, leading us to schedule more breakout sessions in the second part of our un-workshop. According to our exit survey, 70\% of our participants were satisfied with the proposed timeline.

\section{Technology platforms}\label{platformsparagraph}

\subsection{Considerations before choosing a technology platform}
\paragraph{Robustness.} One of the primary concerns is robustness. With online events of an even moderately large scale, a lot of platforms can cut out or fail, and getting attendees to rejoin a meeting once that has happened can be difficult. A good platform should be fairly robust against calls cutting out, network glitches, etc. (assuming the host’s internet connection is stable, has high throughput in Mbps, and is consistent). 

\paragraph{Ease of use.} Ensure that the platform chosen for your event is accessible and easy to understand for the participants, while avoiding the need for overly-difficult software drivers, or proprietary hardware, etc. An important aspect to consider is if it is likely that the participants are familiar with the potential platform to use (e.g. if the same platform was used in previous conferences).

\paragraph{Type of the event (panel, pre-recorded presentations, live Q\&A, poster session).}
All our invited talks were prerecorded using SlidesLive (see \S~\ref{par:slideslive}) and each was followed by 5 minute live question/answering. The transition between pre-recorded and live Q\&A was smooth thanks to the integration  between SlidesLive and Zoom. 
Thanks to platforms like Zoom, streaming video as part of a live event has gotten dramatically easier, but we strongly recommend virtual conference organizers to consider a live component in their program to boost attendee engagement. 
To encourage more interaction, we included 27 breakout sessions which were held via Zoom and were fully live. These sections were viewed as extremely helpful by participants, according to our exit survey.  
Additionally, we also provided a 2-hour live poster session in Gather.town where poster presenters and attendees could meet and talk around their posters; this has been highlighted as the most fun part of our un-workshop and got a lot of attention. 
\paragraph{Length (hours, days).}
We decided to hold the event on the first day of the conference, similar to previous years, which also helps junior researchers to find a conference-buddy during the event as in in-person conferences. The un-workshop lasted for 17 hours on July 13th. We spread the 10-hour program to 17 hours to cover almost all time zones around the world. 
\paragraph{Sessions in parallel.}
We had 27 breakout sessions in total that were spread across 4 time slots. We assigned 4 and 10 parallel streams in different sessions. 
\paragraph{Number of attendees.}
Almost 400 participants filled in our pre-event participation form. However, SlidesLive informed us that our our un-workshop received around 2000 views during the day of the event.  
\paragraph{Cost.}
While choosing the platform, make sure of the plans and pricing you need to accommodate participants. For example, the standard plan of Gather.Town only covers up to 500 people at the time whereas we needed 1000. So, we had to subscribe for the premium plan. Also, for breakout sessions we had to enable the breakout function in Zoom meetings and this was only designed for more expensive plans. Therefore, we strongly recommend that organizers estimate their needs, check the prices of different online platforms from available companies, budget their available funding, and then make a purchase. Zoom enterprise plan hosts up to 500 participants which can be extended to 1000 by using optional add-on plans for large meetings. 
\subsection{Potential broadcast platforms}
Some of the most popular broadcast platforms are discussed in the following.
\paragraph{Zoom.} This works well for webinars or breakout rooms. There are several options that have differing costs, number of allowable attendees, and tools (chat, video for attendees, poll). Overall, this seems to be a more robust option that can handle a large number of participants. It can accommodate screen sharing, either by the host or by a designated participant. The whiteboard feature is generally useful, and can allow for multiple participants to contribute at the same time, but there is some lag and drawing and note taking features are very basic at times.

Zoom has the distinct advantage of being the ``lingua franca'' of remote conferencing - many people are familiar with it, have the drivers and additional software installed already, etc. We were able to ask breakout session organizers to use their own personal Zoom links if they happened to have the paid version of Zoom (often provided by universities and employers). This saved a lot of headaches, as we were able to outsource some of the ``stage management'' to the breakout session leads. 

\paragraph{SlidesLive.}\label{par:slideslive} 
This was provided to us by the ICML conference organizers. We were required to upload pre-recorded videos well in advance of the event to allow time for video quality conversion and quality control, which was provided by SlidesLive. 

\paragraph{Sli.do} Sli.do is principally a platform for audience members to ask questions to a panel. It allows for upvoting of questions by other audience members. This is a very smooth interface, and provides some advantages over ``traditional'' in-person conference queues for asking questions, or working off of questions asked in a chat interface on, e.g., Zoom. It can be much less intimidating to propose a question via Sli.do, watch others +1 your question, and then have the question posed to the panel by a moderator, as compared to asking panelists directly. The upvote feature also allows for more democratic prioritization of questions. Sli.do offers a polling option as well to increase participant engagement, which seems promising, though we didn’t thoroughly test it.

\paragraph{Gather.town.} Gather.town is a virtual gathering space where you can host an online conference, create your customized virtual office, hold a social with your friends and watch a movie together etc. Check their website (https://gather.town/) for more information 


\paragraph{Crowdcast.} An in-browser platform that is inexpensive, accommodates a very large audience, is easy to handle, allows for parallel sessions with only one link, and makes recordings directly available. The “Ask a question box” is pretty useful. It is also accessible via Zoom webinar – which makes the Q\&A sessions easier to handle and more productive than with a chat only. It also offers a green room, which allows organizers and invited speakers to check the setup before going live or while broadcasting live.

\paragraph{YouTube Live/Facebook Live.} While it is a familiar platform that can be paired with Zoom, this is better suited for more casual presentations. It might be better suited to situations where audience interaction can be less information-dense (e.g. general reactions as are available on Facebook), questions can be queued more easily and are much more ephemeral - generally, the chat feature does not allow for deep connection between attendees. There is also limited user-verification - who’s to say, e.g., YoshuaBengio on YouTube is actually the Yoshua Bengio? Similarly, many use their personal email accounts for YouTube and do not have their professional emails linked to YouTube - maybe a prominent researchers’ YouTube handle is ~*~SuperCoolFortniteStreamer~*~. 

\subsection{Final choices}

\paragraph{Breakout sessions.}
For the breakout sessions, we used Zoom meetings and enabled the breakout room functionality. An alternative to Zoom could be Hopin. It is much easier than Zoom since the organizers do not require to collect the Zoom links and passwords. They can just create an event in Hopin and create multiple parallel sessions. 
Also, Zoom provides whiteboard functionality which is pretty helpful for breakout sessions. If collecting Zoom links is not an issue, we would still suggest using that. 

\paragraph{Poster session.}
For poster sessions, we used the Gather.town platform. Gather.town was a very joyful environment for both presenters and attendees. It was very similar to in-person poster sessions and having real human interactions. However, we had some issues in uploading posters and links to their GUI. Also, many participants reported that they couldn't get inside the poster session even though we had capacity for more attendees.
Poster sessions on Twitter can also be interesting as it encourages engagement between people who are not registered for the conference. 
\paragraph{Panel.}
For our panel session, we relied on the broadcasting/streaming infrastructure provided via SlidesLive paired with the Zoom conference call functionality we were well versed with. All participants in the panel were able to hop on pretty easily and frictionlessly.
\paragraph{Invited Talks and Q\&A.}
For our invited talks, we used the Sli.do platform.
We had a moderator on-call to screen the questions and select the most popular ones and re-direct them to an organizer to ask the speaker. 

\section{Dry runs}
We held two different types of dry run, one for breakout session leaders/facilitators over Zoom and another one for poster presenters at Gather.town platform. 
\subsection{Breakout Session Leaders and Facilitators} 
We recommend all leaders and facilitators arrive in the Zoom link at least 15 minutes before the breakout session starts, then the leader who is host of the Zoom link can make all other leaders and facilitators co-hosts. Leaders and facilitators can use this time to get familiar with the whiteboard, breakout rooms, security toolbar, etc. We also recommend leaders and facilitators of a breakout session stay in contact via a private side channel (e.g. group message), turn off desktop notifications during the breakout session and mute when they are not speaking.
\subsection{Gather.town Poster Presenters}
We asked all the presenters to: 
\begin{enumerate}
    \item Select the right avatar for poster presenters
    \item Test Gather.town experience -- walk around, view someone else’s poster, etc.
    \item Test Gather.town controls -- how to change video quality. 
    \item Test the video, audio, internet connection (see Gather.town Technology Requirements to troubleshoot). 
    \item Test the interaction distance (see Control toolbar) to find an interaction distance that is comfortable for you.
\end{enumerate}
\newpage
\paragraph{Here is the general dry run checklist for all speakers and participants}
\begin{enumerate}
    \item Internet speed
    \item Audio quality and video quality (including position of camera and lighting)
    \item Familiarity with the platform used, i.e. sharing screen etc.
\end{enumerate}

\section{Day-of-Event}
\subsection{Schedule of activities} As previously mentioned, we arranged the conference so that there were two different time blocks - one block of talks and breakout sessions that were compatible with participants in Europe, Africa, and parts of Asia, a significantly long break, and a second time block with better compatibility for those in North America. The organizers worked in shifts to serve as organizer-on-call. The organizer-on-call for a given shift was tasked with hosting the main Zoom room (where the conference talks were taking place), serve as point of contact for supervolunteers, and to act as a liason for the live panelists when necessary. 
\subsection{Volunteers}
\paragraph{Volunteer structure} Due to the wide range of time zones that our event covered, we implemented a structure of shifts where ``supervolunteers'' (those who have volunteered for WiML in the past, or have a particularly deep knowledge of Zoom), who reported to an organizer, supervised all other volunteers. This was to allow reasonable waking hours and increase the available bandwidth for the organizers - for example, the first shift began at 9AM CET, which is very early in the morning for those in, e.g., PST timezones. Generally,
we divided up the responsibilities among the different volunteers based on tasks:
\begin{enumerate}
    \item \emph{Moderators}: Check all communication channels, including Slack, Zoom, and Rocket.chat, for violations of the Code of Conduct. Moderator volunteers were asked to familiarize themselves with the Code of Conduct.
    \item \emph{Tech Support}: Help with issues related to getting Zoom / Slack / Rocket.chat / Gather.town up and running, troubleshoot minor technical details, etc. Mainly answered questions posed in a dedicated Rocket.chat channel that was accessible to all attendees. These volunteers were referred to a handful of internal documents to use as guides during their shifts.
    \item \emph{General Infodesk}: Guide participants regarding general organizational questions that would not fall under the jurisdiction of ``tech support''. Mainly, they answered questions posed via a dedicated public Slack channel. These volunteers were asked to make themselves very familiar with the schedule of the conference.
\end{enumerate}
We requested that volunteers check in with their supervolunteer supervisor in their team's dedicated private Slack channel (within the Slack team intended for all attendees) that was only visible to volunteers and supervolunteers of the same assignment (e.g. \#ModeratorVolunteers, \#TechSupportVolunteers, etc). We provided volunteers with access to a handful of documents designed to cover anticipated questions, and asked that they familiarize themselves with these documents and use them as reference during their shifts. Creating volunteer documentation in an FAQ format resulted in surprisingly few questions from confused volunteers on the day of the event. 
While participants were encouraged to use Rocket.chat to ask technical support questions, we saw a lot of participants use the general-information Slack channel to ask technical support questions regardless. Other organizers should perhaps anticipate that participants who need help at an online event might ask in channels outside of a designated ``help line''. 

\newpage
\section{Communication internal and external}

As in any collaborative project, communication is the most 
vital part to ensuring effective teamwork.
This is particularly true when organizing a virtual event over a relatively short period of time.

\subsection{Planning stage}
In the introductory meeting with all the organizers, a weekly meeting 
was fixed that worked with everyone's schedule.
Particular care should be taken with considering the timezone of all members.
Meeting times early in the morning of one member might result in that member not being up to speed with the latest developments on the project; meeting times late at night for another member can prevent said member from participating in tasks that is decided on during the meeting and are easy to solve right after the meeting, with the whole team being updated and together. 
With members from central Europe, West- and East-coast North America, all
of our meetings had to be held between 5pm and 10pm GMT/UTC.
Each organizer had a task domain, so each task could always easily be assigned 
to one person responsible overseeing the task.
\paragraph{Weekly meetings} The meeting schedule consisted of weekly Zoom meetings including all five organizing members. Approaching the day-of-event, additional meetings were called for subteams to work on specific tasks.
From our experiences we want to emphasize the importance of ``overcommunicating'' during these meetings. 
With members spread out over several timezones, it can easily occur that 
member have asynchronous levels of information, the task assignment is not clear etc.
Short recaps of what the current state is and next steps are in every tasks can prevent misunderstandings and delays.

We and the other organizers of the un-workshop established a Slack channel to communicate with each other and with members of the WiML Board of Directors.

All files, spreadsheets, drafts of publications (emails$/$Twitter$/$website) were shared in a shared Google Drive accessible to all organizers and WiML board members overseeing the organizers. 
Having a shared drive is highly recommended for projects with a high level 
of bureaucracy, shared spreadsheets etc., to avoid these documents being cluttered on a Slack channel and hard to manage at a later point in time.

\subsection{Day-of-event (DOE)}
On the day-of-event, we anticipated numerous inquiries from participants 
regarding e.g. problems with the technology platforms used and questions 
regarding the program and organization of the event.
In order to keep the organizers free to handle any emergencies during the event, we implemented a hierarchical volunteer structure to address incoming
questions from attendees.
In this structure, incoming questions are directed at volunteers who 
worked in 1-hour shifts and were recruited from the attendees who 
indicated interest in their sign-up form.
We paid attention to the volunteers being familiar with the technology 
platforms used.
Above them, nine supervolunteers were responsible for handling questions
from volunteers and organizing check-in and -out of volunteers for their shift.
The supervolunteers were recruited from attendees experienced in organizing similar events.
Only if supervolunteers could not solve a particular request, it was directed to the un-workshop organizers.

The platform used for particular communications were:
\paragraph{Attendees - Organizers/Volunteers.} Attendees who watched the live stream online could use the Rocket.chat feature on the \emph{icml.cc} website
to give immediate feedback on the event or ask questions to volunteers.
\paragraph{Volunteers - Supervolunteers/Organizers.} For the DOE we opened an additional Slack (external) workspace to coordinate volunteers and provide a channel where volunteers' questions could be answered by supervolunteers, volunteers could check-in and check-out in private channels, and supervolunteers and un-workshop organizers could communicate with one another. This Slack workspace was also available to general attendees, albeit with restricted access to volunteer-only channels.
\paragraph{Organizers - Organizers.}  For internal communication in the organizing team, the internal Slack channel was used and phone numbers were exchanged for urgent matters.
\section{Evaluation and Postmortem}
We sent a survey to all participants who signed up for the un-workshop to gather feedback on their experience.
Specifically, we asked for feedback on each program item, and if it should be repeated in the way it was held. 
\paragraph{Exit survey results.}
The talks and panel discussion received good ratings and -- unsurprisingly, as their format was most similar to the usual in-person conference format -- were liked by attendees as-is.
For the Poster Session, several people reported that they did not attend, 
presumably due to the late timing in Europe or the technical issues with 
the platform, but feedback was positive from the participants who attended.
Breakout sessions received very positive feedback, with more than 95\% 
responses recommending to repeat them as they were held or only with 
minor changes.
The feedback for the split schedule was positive, with three-quarters of responses indicating that the timezone worked for them.

\paragraph{Potential improvements.}
We came up with several potential improvements for future virtual events. 
Firstly, we used many different platforms to provide the optimal technology for each program item.
This lead to some confusion as how to access the different channels, 
i.e. breakout session Zoom rooms, livestream on \emph{icml.cc}, pre-recorded 
videos of posters etc., 
One improvement could be to provide a guide, ideally as a video,
to demonstrate the navigation through the different program points 
and how to access each program element.

Other improvements could be to check for capacity issues within the chosen platform(s) for activities with many simultaneous interactions of participants, i.e. the poster session.
Scheduling different groups to interact, or having several poster sessions 
at different timepoints could prevent issues regarding capacity.

\section{Diversity and inclusion}

\paragraph{A global conference.} One of the unintended benefits of ICML switching to a virtual conference format was the increase in participants who might have otherwise been unable to attend. This had ramifications for how our scholarship funding was used and the composition of our workshop attendees, among other factors. 
\subsection{Funding} 

This year, due to the virtual form of events, WiML could offer registration fee funding (to every person who applied, and met the criteria for receiving funding). Online conferences tend to be many times cheaper than in-person conferences, both in terms of covering registration as well as in terms of not having to pay for travel/lodging.

Reimbursement directly through ICML required an arrangement between WiML and the ICML Organizers. Other conference organizers might need to negotiate this payment arrangement. This is also significant as we received a lot of applications for funding coverage from people in developing economies, where the registration fee in virtual conferences is considered a large amount of money that would require a loan or credit line to be extended, etc. Thus, having a registration fee funding system where attendees pay the fee, and are then later reimbursed for the cost might result in some scholarship recipients not attending at all. Other online conference organizers might want to try to negotiate reimbursement before the conference taking place, to avoid the requirement of attendees paying the fee upfront.

\paragraph{Passlist.} In addition to ``approving'' attendees through a passlist to avoid unwelcome interlopers, one might also want to check the list of words that are blocked from use or are otherwise penalized. For example, some often-misused (but in this case, necessary and applicable) words were banned on one of the platforms we used during our joint poster session with other affinity groups, Queer in AI and Latinx in AI. 

\subsection{Safety as it pertains to special interest group workshops}
A virtual environment brings with it a number of challenges related to anonymity and appropriate behavior, some intentional, some unintentional.

\paragraph{Code of Conduct (CoC).}
We recommend using a registration page as an excellent opportunity to ensure that your attendees have read the Code of Conduct for your conference, if your organization has a code of conduct published.
We required the WiML Code of Conduct (https://wimlworkshop.org/conduct/) to be followed by all participants, organizers, speakers, presenters, and sponsors of WiML ICML Un-Workshop; confirmation that a participant would adhere to the code was required in order to register for the event. Organizers of a conference have to decide on the code of conduct statements early on in the planning process, as this is a very important aspect of virtual events and allows safety and privacy to be preserved during the online events.
Inappropriate or unprofessional behavior that interfered with another participants' full participation was not to be tolerated at the event. Any violation of the CoC was to be reported through email to the organizers and board of directors. Fortunately, we encountered no CoC infractions. 

\paragraph{Avoiding Zoombombing.}
With the advent of more and more remote conferences relying on Zoom, a handful of issues have come to the forefront. Due to these issues, some companies avoided using Zoom to avoid ``zoombombing''.  
The threat of malicious attacks is especially heightened given that attacks of this type in an online environment often disproportionately target women and people of color (see Gamergate, \href{https://twitter.com/AnimaAnandkumar/status/1298400611549868033}{Anima Anandkumar's tweet} regarding getting zoom-bombed at the Knowledge Discovery in Databases (KDD) conference) and what to do in case that happens (see \href{https://twitter.com/wiebketous/status/1298410864903630851}{Wiebke Toussaint's tweet}). 

\paragraph{Suggested Actions To Take.}
As an organizer, make sure to keep Zoom links private and difficult to guess, perhaps accessible only behind a password-protection wall. Also, explicitly communicate to attendees that Zoom links should not be shared with those outside of the conference and emphasize that they should use password protected Zoom links.

\section{Security and privacy matters} 
\paragraph{Full Name Policy.}
Across all platforms, we encouraged participants to use their real full name rather than their nicknames, Twitter handles, etc. This was to deter bad actors from attending the event while pretending to be someone else, and then possibly breaking the code of conduct. Implementing a ``full name'' policy might also have advantages in follow-ups after the conference\,- it is easier to find and contact someone afterwards by searching according to name versus once-used Slack handle.

\paragraph{Chatham House Rules.} While typically more common in policy conversations, it may be helpful to consider applying Chatham House Rules to your breakout sessions. This effectively means that what is said can be on record, however, it will not be attributed to a specific attendee. This may be helpful for accommodating people who refuse to be on record due to privacy concerns. 

\paragraph{Security considerations for Gather.town.}
To ensure all poster presenters and participants are identified, we asked everyone to only use their real full name in all platforms, including those used for the poster session, namely Zoom and Gather.town. We defined a visual code for particular roles in Gather.town, e.g. all volunteers in Gather.town should have chosen an avatar with a red t-shirt, organizers were identified by their avatars wearing a green t-shirt, and diversity and inclusion (D \& I) chairs' avatars wore blue ones. 
Access to Gather.town was gated to only ICML registered attendees. We did encounter some issues with those who registered at the last minute for ICML. If there is some way to dynamically update a passlist on Gather.town in the future, this would be a major improvement.

\newpage
\section{Common failure modes in Gather.town} 
Before the un-workshop, we communicated the following to participants so that they could be prepared in case of any issues during the event. 

\paragraph{Permission problem to access video and microphone.} Upon entering Gather.town, you will be asked to give Gather.town permission to access your video and microphone. If you do not provide permission or block the pop-up, you will see this message: ``We need your video or audio permission for Gather.town to run properly''. Read Gather.town instructions (https://gather.town/video-issues) for how to fix this.
\paragraph{Audio/Video issues.} There are several reasons this could be happening.
    Each user can see a maximum of eight other users’ videos at once, other users can use audio-only (can only be heard, not seen). If a user cannot see more than a few videos, this might be one reason.
    Another possible reason is that another application is capturing the microphone or camera. For example, Zoom on Windows takes full control of the camera. Close any other apps that might use the microphone or camera. 
    Yet another reason is that if the internet connection is not very good, video may be cut automatically, leaving only audio. 
    You can lower your Gather.town video quality and close any other programs that might be using Internet bandwidth. Also, consider moving to a better location in your room to see if you can get better WiFi.
\paragraph{Physically blocked, cannot move.} If you are blocked on all four sides and cannot move, a pop-up should appear asking if you want to move away. Press the key for ``yes'' so that your avatar gets ``transported''. If you are blocked and a pop-up does not show up, ask for help in the public chat.
\paragraph{Wrong camera/microphone input.} Read Gather.town instructions (https://gather.town/video-issues) on how to switch your camera/microphone input.

\section{Conclusions} 
Despite the organizational challenges, the un-workshop program was carried out with no disruptions, and most importantly, with great engagement and participation from our attendees. Our attendees had the opportunity to: (i) learn and discuss scientific and technical developments in machine learning, (ii) meet and discuss with talented researchers, and (iii) promote some of the work done by women and non-binary persons working in our field. The virtual setting, reduced price of the ICML conference, and WiML funding to fund ICML conference registration fees for un-workshop attendees all likely contributed to high attendance at the un-workshop. Based on our post-event surveys, for many of our attendees, this was their first-ever machine learning conference, and a good opportunity to meet with colleagues interested in similar topics, as well as promote their work. We are heartened by this success. 
%
%

\section*{Acknowledgements}
The authors would like to thank Amy Zhang, Sarah Tan, Jennifer Healey, Ti John, and Nils Murrugarra for their support and feedback on this or earlier versions of the report. The WiML Virtual Un-Workshop could not have happened without contributions from sponsors, volunteers, and many others as acknowledged in https://wimlworkshop.org/icml2020/committee/, as well as collaboration with affinity groups such as Queer in AI and LatinX in AI.

\section*{Disclaimer}
The views expressed in this document are solely those of the authors and do not represent the opinions of any entity whatsoever with which they ever have been, are now or in the future will be affiliated. In particular, this document does not express the opinion of WiML, WiML Board of Directors, WiML sponsors, or ICML. The authors of this document were recruited to help organize the WiML event at ICML, which this year was held as an ICML workshop titled  \emph{"WiML Un-Workshop”}. For the purposes of disambiguation, this event was not \emph{“WiML Workshop”} which is a separate event co-located with NeurIPS.

\bibliographystyle{abbrvnat}
\bibliography{references}

\end{document}